\documentclass{jnmp}

\usepackage{amsmath}

\JNMPnumberwithin{equation}{section}

\newtheorem{theorem}{Theorem}[section]
\newtheorem{lemma}{Lemma}[section]
\newtheorem{proposition}{Proposition}[section]

\theoremstyle{definition}
\newtheorem{definition}{Definition}[section]
\newtheorem{remark}{Remark}[section]
\newtheorem*{example}{Example} 

\begin{document}

%
\renewcommand{\evenhead}{P Guha}
\renewcommand{\oddhead}{Volume Preserving Multidimensional Integrable Systems}

%
\thispagestyle{empty}

\FirstPageHead{8}{3}{2001}{\pageref{guha-firstpage}--\pageref{guha-lastpage}}{Article}

\copyrightnote{2001}{P Guha}

\Name{Volume Preserving Multidimensional Integrable Systems
and Nambu--Poisson Geometry}
\label{guha-firstpage}

\Author{Partha GUHA}

\Address{S.N. Bose National Centre for Basic Sciences, JD Block, Sector-3, Salt Lake \\
Calcutta-700098, India}

\Date{Received June 14, 2000; First Revision February 9, 2001;
 Second Revision March 12, 2001;
Accepted March 24, 2001}

\begin{abstract}
\noindent
In this paper we study generalized classes of volume preserving
multidimensional integrable systems via Nambu--Poisson mechanics.
These integrable systems belong to the same class of dispersionless
KP type equation. Hence they bear a close resemblance
to the self dual Einstein equation. All these dispersionless
KP and dToda type equations can be studied via twistor
geometry, by using the method of Gindikin's pencil of two forms.
Following this approach we study the twistor construction of
our volume preserving systems.
\end{abstract}

\rightline{\bfseries \itshape Dedicated to the memory of Dr. B~C~Guha}

\section{Introduction}

In this article we study volume preserving diffeomorphic integrable
hierarchy of three flows~[1]. This is different from the usual two flows
cases, and this can be studied via Nambu-Poisson geometry. It is already known
that a group of volume preserving diffeomorphisms in three dimension plays a
key role in an Einstein--Maxwell theory where the Weyl curvature is self-dual and
the Maxwell tensor is algebraically anti self-dual. Later Takasaki~[2]
explicitly showed how volume preserving diffeomorphisms arises in integrable
deformations of self-dual gravity.

Nambu mechanics is a generalization of classical Hamiltonian mechanics,
introduced by Yoichiro Nambu~[3]. At the begining he wanted to formulate a
statistical mechanics on ${\mathbb R}^3$, emphasizing that the
only feature of Hamiltonian mechanics one should preserve
is the Liouville theorem. He considered the following equations of motion
\[
 \frac{d{\bf r}}{dt} = \nabla u({\bf r})  \wedge
\nabla v({\bf r}), \qquad {\bf r}= (x,y,z) \in {\mathbb R},
\]
where $x$, $y$, $z$ are dynamical variables and $u$, $v$ are two functions of
 ${\bf r}$. Then Liouville theorem follows from the identity
\[
\nabla \cdot (\nabla u({\bf r}) \wedge \nabla v({\bf r}))= 0.
\]

He further observed from above equation that the
evolution of a function on ${\mathbb R}^3$ is given by
\[
\frac{df}{dt} = \frac{\partial (f,u,v)}{\partial (x,y,z)},
\]
where the right hand side is the Jacobian of the mapping ${\mathbb R}^3
\longrightarrow {\mathbb R}^3$ given by
\[
(x,y,z) \longrightarrow (f,u,v).
\]
The right hand side can be interpreted as a generalized Poisson
 bracket.
Hence the binary operation of Poisson bracket of Hamiltonian mechanics is
 generalized to $n$-ary operation in Nambu mechanics. Recently Takhtajan~[4,~5]
 has formulated its basic principles in an invariant geometrical
 form similar to that of Hamiltonian mechanics.

In this paper, we shall use Nambu mechanics to study generalized
volume preserving diffeomorphic integrable hierarchy.
These classes of integrable systems are closely related to the self dual
Einstein equation, dispersionless KP
equations etc. In fact we obtain a higher dimensional analogue of
all these systems. It turns out that all these systems can be
written in the following form:
\[
d\Omega^{(n)}= 0, \qquad  \Omega^{(n)} \wedge \Omega^{(n)} = 0.
\]
For $n=2$ we obtain all the self dual Einstein and dKP type equations.

Hence we obtain a common structure behind all these integrable system, so
there is a consistent and coherent way to describe all these systems.
Here we unify all these classes of integrable systems by Gindikin's pencil
or bundle of forms and Riemann--Hilbert problem (twistor description).
Gindikin introduced these technique to study the geometry of the solution of
self dual Einstein equations. Later Takasaki--Takebe~[6,~7] applied
it to dispersionless KP and Toda equations.

This paper is organized as follows.
In Section~2 we present a tacit introduction of Nambu--Poisson geometry.
In Section~3 we construct our volume preserving integrable systems via
Nambu--Poisson geometry. If one carefully analyse these set of equations,
then one must admit that they bear a close resemblance with the {\sl volume
preserving KP equation}, so far nobody knows about this equation. It is
known that area preserving KP hierarchy (= dispersionless KP hierarchy)
plays an important role in topological minimal models (Landau--Ginzburg
description of the A-type minimal models). So we expect volume preserving
KP hierarchy may play a big role in low dimensional quantum field theories.
Section~4 is dedicated to the twistor construction
of these systems.

\subsection*{Notations}

\begin{enumerate}
\itemsep0mm
\item ${\cal M}$ : Nambu-Poisson manifold.
\item $\eta$ : Nambu-Poisson polyvector (tensor).
\item $J_{\eta}$ : Bundle map associated to $\eta$.
\item ${\cal X}_{f_1 \cdots f_{n-1}}$ : Hamiltonian vector field associated to $\eta$.
\item $L$, $M$, $N$ :
Formal Laurent series, Lax operators involving volume preserving systems.
\item ${\cal L}$ : Lax operator for dispersionless KP hierarchy.
\item ${\cal K}$ : Orlov operator in dispersionless KP.
\item $\Omega$ : two form in dispersionless KP.
\item $\Omega^{(3)}$ : three form in volume preserving integrable systems.
\item ${\cal T}$ : curved twistor space.
\item $(P,Q)$ or $(P^{\prime},Q^{\prime})$  : Pair of Darboux coordinates.
\item $e^{ij}$ : one forms or tetrad.
\end{enumerate}

\section{Nambu--Poisson manifolds}

The modern concept of Nambu--Poisson structure was proposed by Takhtajan in
1994 in order to find an axiomatic formulation for the $n$-bracket operation.
Let ${\cal M}$ denote a smooth $n$-dimensional manifold and
$C^{\infty }({\cal M})$ the algebra of infinitely differentiable
real valued functions on ${\cal M}$. A manifold
${\cal M}$ is called a Nambu--Poisson manifold if there exits a
${\mathbb R}$-multi-linear map
\[\label{guha:1}
\{ ~,\ldots,~ \}~:~[C^{\infty }({\cal M})]^{\otimes n} \rightarrow
C^{\infty }({\cal M}).
\]
This is called Nambu--Poisson bracket of order $n$
 $\forall \; f_1 , f_2 , \ldots , f_{2n-1} \in C^{\infty }({\cal M})$.
This bracket satisfies
\begin{gather}
\{ f_1, \ldots ,f_n \}=(-1)^{\epsilon(\sigma)}\{ f_{\sigma(1)}, \ldots , f_{\sigma(n)} \},\label{guha:2}\\
\{ f_1 f_2, f_3, \ldots ,f_{n+1} \}= f_1 \{f_2, f_3, \ldots , f_{n+1} \} +
\{ f_1, f_3, \ldots, f_{n+1} \} f_2,\label{guha:3}
\end{gather}
and
\begin{gather}
\{ \{ f_1, \ldots , f_{n-1}, f_n \}, f_{n+1}, \ldots, f_{2n-1} \}  \nonumber\\
\qquad {}+ \{ f_n, \{ f_1, \ldots, f_{n-1}, f_{n+1} \}, f_{n+2}, \ldots , f_{2n-1} \} +  \cdots
\nonumber\\
\qquad {}+ \{ f_n, \ldots ,f_{2n-2}, \{ f_1, \ldots , f_{n-1}, f_{2n-1} \}\}\nonumber\\
\qquad{}=  \{ f_1, \ldots , f_{n-1}, \{ f_n, \ldots , f_{2n-1} \}\}, \label{guha:4}
\end{gather}
where $\sigma \in S_n$ --- the symmetric group of $n$ elements --- and
$\epsilon(\sigma)$ is its parity.  Equations~(\ref{guha:2}) and (\ref{guha:3}) are the standard
skew-symmetry and derivation properties found for the ordinary ($n=2$) Poisson
bracket, whereas~(\ref{guha:4}) is a generalization of the Jacobi identity and was called
in~[2] the fundamental identity. When $n=3$ this fundamental identity reduces to
\begin{gather*}
\{\{ f_1, f_2, f_3 \}, f_4, f_5 \} + \{ f_3, \{ f_1, f_2, f_4 \}, f_5 \}\nonumber\\
 \qquad {}
+ \{ f_3, f_4, \{ f_1, f_2, f_5 \} \}   =  \{ f_1, f_2, \{ f_3, f_4, f_5 \} \}. \label{guha:5}
\end{gather*}

\subsection{Hamiltonian geometry}

The Nambu--Poisson bracket is geometrically realized by the
Nambu--Poisson polyvector $\eta \in
\Gamma (\wedge ^n TM)$, a section of $\wedge ^n TM$, such that
\[
\{ f_1,\ldots,f_n \} = \eta (df_1,\ldots,df_n),\label{guha:6}
\]
in local coordinates $(x_1,\ldots,x_n)$ this is given by
\[
 \eta = \eta_{i_1 \ldots i_n}(x) \frac{\partial}{\partial x_{i_1}}
\wedge \cdots
\wedge \frac{\partial}{\partial x_{i_n}},\label{guha:7}
\]
where summation over repeated indices are assumed.

The structure is regular if $\eta \neq 0$. The canonical example of a
Nambu--Poisson structure of order $ n \geq 2$ is the one induced by a volume
form on an oriented manifold
\[
\{f_1, \ldots ,f_n \} = \eta (df_1, \ldots ,df_n ).
\]

The Nambu--Poisson polyvector defines a bundle map
\[
J_{\eta}~:~ \underbrace{T^{\ast}{\cal M} \times \cdots T^{\ast}{\cal M}}_{n-1}
\longrightarrow T{\cal M}\label{guha:8}
\]
given by
\[
\langle\beta, J_{\eta}(\alpha_1, \ldots ,\alpha_{n-1})\rangle
= \eta (\alpha_1, \ldots ,\alpha_{n-1}, \beta),
\]
where
$\beta, \alpha_i \in \Omega^1(M)$.

The Hamiltonian vector field is defined by
\[
{\cal X}_{f_1, \ldots ,f_{n-1}} = J_{\eta}(df_1, \ldots ,df_{n-1}).\label{guha:9}
\]
This is a Hamiltonian vector field of $(n-1)$ functions. Thus, we
can express the fundamental identity as
\[
L_{{\cal X}_{f_1 \cdots f_{n-1}}}\eta (dg_1 \cdots dg_n) = 0,\label{guha:10}
\]
where $L$ is the Lie derivative. This shows that the Hamiltonian
vector fields are the infinitesimal automorphism of the Nambu--Poisson tensor.

The polyvector field $\eta \in \Gamma(\wedge^n TM)$ defines an $n$-ary
Poisson bracket if and only if either $n$ is even and the Schouten bracket
satisfies $[\eta , \eta] = 0$, or $n$ is odd if $\eta$ satisfies
the conditions (see for detail [8])
\begin{gather*}
 i_{\alpha}(\eta) \wedge i_{\beta}(\eta) = 0, \quad
\forall \; \alpha, \beta \in T^{\ast}{\cal M},\qquad
\sum_{j=1}^{n} (i_{dx_j}\eta) \wedge \left(L_{\frac{\partial}{\partial x_j}}\eta\right)= 0.
\end{gather*}

In this framework we can introduce the bihamiltonian structure also.
A bihamiltonian system is prescribed by specifying two sets
of Hamiltonian function $ H = H_1, \ldots ,H_n$ and $ h = h_1, \ldots, h_n$,
where $ H_i, h_i \in C^{\infty}({\cal M})$,
\[
 {\cal X} = J_{\eta}^{0}(dh_1, \ldots ,dh_n) =
J_{\eta}^{1}(dH_1, \ldots ,dH_n),\label{guha:11}
\]
where $J_{\eta}^{i}$s are the bundle maps.

\pagebreak

One more interesting algebraic structure pops up in the
Nambu--Poisson geometry~--- this is called Leibniz algebra.
The Nambu--Poisson polyvector $\eta$ induces a homomorphism
of vector bundles
\[
J_{\eta}^{k}~:~ \wedge^k (T^{\ast}{\cal M}) \longrightarrow \wedge^{n-k}(TM),
\]
such that $ J_{\eta}^{k}(\beta) = i_{\beta}\eta$ for $ \beta \in
\wedge^k T_{x}^{\ast}{\cal M}$.
 All these higher covectors form
Leibniz algebra.
It was shown by Ibanez et.~al.~[9] that the bracket of $(n-1)$
 forms on ${\cal M}$ is the ${\mathbb R}$-bilinear operations
\[
[\cdot , \cdot ]_{\rm Leibniz} ~:~ C^{\infty}(\wedge^{n-1}T^{\ast}{\cal M})
\times C^{\infty}(\wedge^{n-1}T^{\ast}{\cal M}) \longrightarrow
C^{\infty}(\wedge^{n-1}T^{\ast}{\cal M}),
\]
given by
\[
 [\alpha , \beta ] = L_{J^{\alpha}}\beta + (-1)^n (i(d\alpha)\eta)\beta,
\]
for all $\alpha, \beta \in C^{\infty} (\wedge^{n-1}T^{\ast}{\cal M})$.

Any Leibniz algebra (${\cal A}, \circ)$ satisfies
\[
 f \circ ( g \circ h) = ( f \circ g) \circ h + g \circ ( f \circ h), \qquad
f,g,h \in C^{\infty}({\cal M}).
\]

A morphism in the Nambu--Poisson category is a map
\[
 \phi ~:~ M \longrightarrow N
\]
between Nambu--Poisson manifolds preserving Nambu--Poisson
brackets:
\[
\{f_1, \cdots ,f_n \}_N \circ \phi =
 \{f_1 \circ \phi, \cdots , f_n \circ \phi \}_M.
\]

The fundamental difference between the Nambu--Poisson bracket
and the classical Poisson case is that for $n \geq 3$ the Nambu--Poisson
polyvector $\eta$ is decomposable, i.e. it has rank $n$ at points where
it does not vanish.

The Nambu--Poisson polyvectors were charcterized by Takhtajan. The polyvector
field~$\eta$ is a Nambu--Poisson tensor if and only if the natural component of~$\eta$
satisfy certain conditions.
It was stated in [4] that the fundamental identity (\ref{guha:4}) is equivalent to
the following algebraic and differential constraints on the Nambu--Poisson tensor~$\eta$:
\[
{\rm (A)}   \quad S_{ij} + P(S)_{ij} = 0,\label{guha:12}
\]
for all multi-indices $i=\{i_1,\ldots,i_n \}$ and
$j=\{j_1,\ldots,j_n \}$ from the set $\{1,\ldots,N \}$,
where
\begin{gather*}
S_{ij} = \eta _{i_1 \ldots i_n} \eta _{j_1 \ldots j_n}
+ \eta _{j_n i_1 i_3 \ldots i_n } \eta _{j_1 \ldots j_{n-1} i_2}
+ \cdots\nonumber\\
\qquad {}+ \eta _{j_n i_2 \ldots i_{n-1} i_1 } \eta _{j_1 \ldots j_{n-1} i_n}
- \eta _{j_n i_2 \ldots i_n } \eta _{j_1 \ldots j_{n-1} i_1}; \label{guha:13}
\end{gather*}

(B) $P$ is the permutation operator which interchanges the indices $i_1$ and
$j_1$ of $2n$-tensor~$S$, and
\begin{gather*}
 \sum_{l=1}^{N}  \left ( \eta_{l i_2 \ldots i_n}
\frac{\partial \eta_{j_1 \ldots j_n}}
{\partial x_l} + \eta_{j_n l i_3 \ldots i_n}\frac{\partial \eta_{j_1 \ldots
j_{n-1} i_2}}{\partial x_l} + \cdots + \eta_{j_n i_2 \ldots i_{n-1}l}
\frac{\partial\eta_{j_1 \ldots j_{n-1} i_n}}{\partial x_l}
\right) \nonumber \\
\qquad{} =  \sum_{l=1}^{N}  \eta_{j_1 j_2 \ldots j_{n-1} l}
\frac{\partial \eta_{j_n i_2  \ldots i_n}} {\partial x_l},\label{guha:14}
\end{gather*}
for all $i_2,\ldots,i_n,j_1,\ldots,j_n = 1,\ldots,N$.

\pagebreak

It was conjectured in [2] that the equation $S_{ij}=0$ is equivalent
to the condition that $n$-tensor $\eta$ is decomposable,
recently this has been proved by Alekseevsky and author~[10],
Since then several other proofs appeared this
surprising result (see [11, 12] and references therein).

The algebraic identity is the special feature for the Nambu--Poisson structure.
It appears solely in the Nambu--Poisson geometry --- there is no counterpart of this
structure in the usual Poisson geometry.

The following theorem describes the local structure of the Nambu--Poisson
structure of order $n$ with $n \geq 3$.

\begin{theorem}[Local Triviality Theorem]\label{theorem:1}
Let $\eta$ be a Nambu--Poisson polyvector of order $n \geq 3$.
Near any point at which $\eta$ does not vanish there are local
coordinates $x_1, \ldots ,x_n$ such that
\[ \eta = \frac{\partial}{\partial{x_1}} \wedge \frac{\partial}{\partial{x_2}}
 \wedge \cdots \wedge
\frac{\partial}{\partial{x_n}}.
\]
\end{theorem}

Let $\{h_1, \ldots ,h_{n-1} \}$ be the set of Hamiltonian functions,
 and ${\cal X}_{h_1\cdots h_{n-1}}$ be corresponding the Hamiltonian
vector field. Then the equation for the integral curves of
${\cal X}_{h_1 \cdots h_{n-1}}$ is
\[
\dot{x} = J_{\eta}(dh_1, \ldots, dh_{n-1}),\label{guha:15}
\]
 Hamilton's equations of motion.

\subsection{Nambu--Poisson dynamics}

It is also shown in [2] that Nambu dynamics on a Nambu--Poisson phase
space involves $n-1$ so-called Nambu--Hamiltonians $H_1, \ldots, H_{n-1} \in
C^{\infty }({\cal M})$ and is governed by the following equations of motion
\[
\frac {df}{dt} =  \{ f , H_1 ,\ldots,H_{n-1} \}, \qquad \forall \; f \in
C^{\infty }({\cal M}).\label{guha:16}
\]

A solution to the Nambu--Hamilton equations of motion produces an evolution
operator $U_t$ which by virtue of the fundamental identity preserves
the Nambu bracket structure on $C^{\infty }(M)$.

\begin{definition}\label{definition:1}
$ f \in C^{\infty}({\cal M})$ is called an integral of motion for the system if it satisfies
\[
 \{ f,H_1,H_2, \ldots ,H_{n-1} \} = 0.
\]
\end{definition}

\begin{example}
 Let us illustrate how Nambu--Poisson mechanics works in practise.
The example is the motion of a rigid body with a torque about the
major axis introduced by Bloch and Marsden~[13].

Euler's equation for the rigid body with a single torque $u$ about its major
axis is given by
\begin{gather*} {\dot m}_1 = a_1 m_2 m_3, \qquad 
{\dot m}_2= a_2 m_1 m_3, \qquad 
{\dot m}_3= a_3 m_1 m_2 + u, \label{guha:19}
 \end{gather*}
where $ u = -km_1 m_2$ is the feedback,
$ a_1 = \frac{1}{I_2} - \frac{1}{I_3}$,
$ a_2 = \frac{1}{I_3} - \frac{1}{I_1}$ and
$ a_3 = \frac{1}{I_1} - \frac{1}{I_2}$. We assume $ I_1 < I_2 < I_3$.

These equations can be easily recast into generalized
Nambu--Hamiltonian equations of motion
\[
\frac{dm_i}{dt} = \{ H_1, H_2, m_i \}, \label{guha:20}
\]
where the right hand side is given by
\[
\{H_1, H_2, m_i \} := \frac{\partial(H_1, H_2, m_i)}
{\partial(m_1, m_2, m_3)}.
\]

These equations involve two Hamiltonians and these are
\[
  H_1 = \frac{1}{2} \left( a_2 m_{1}^2 - a_1 m_{2}^2\right), \qquad
 H_2 = \frac{1}{2} \left( \frac{a_3 - k}{a_1}m_{1}^2 - m_{3}^2\right).
\]

When $a_1 = a_2= a_3 = 1$ and $ u_3=0$, these set of equations
reduce to a famous Euler equation or Nahm's equation
\[
\frac{dT_i}{dt} = \epsilon_{ijk} [T_j,T_k], \qquad i,j,k = 1,2,3,
\]
where $T_is$ are $SU(2)$ generators.

It is well known that the system of $N$ vortices can be
described by the following system of differential equations~[14]
\[
{\dot z}_n = i\sum_{m\neq n}^{N}
\frac{\Gamma_m}{z_{n}^{\ast} - z_{m}^{\ast}},
\]
where $z_n = x_n + iy_n$ are complex coordinate of the centre of $n$-th
vortex.

The equation of motion of a system of three vortices can be put
in the following form~[15]
\[
{\dot M}_i = \epsilon_{ijk}\Gamma_i M_i(M_j - M_k), \qquad
i,j,k = 1,2,3
\]
where $M_1 = |z_2 - z_3|^{2}$, $M_2 = |z_3 - z_1|^2$
and $M_3 = |z_1 - z_2|^2$.
Again, these equations can be described by Nambu--Poisson geometry.
In fact, there is a large scope to apply Nambu--Poisson formulation
in fluid mechanics [16].
\end{example}

\section{Volume preserving integrable systems}

 In this section we shall follow the approach of Takasaki--Takebe's [6, 7]
method of  area preserving diffeomorphic (or sDiff$(2)$)
KP equation. In fact, they adopted their method from self dual
vacuum Einstein equation theory. In the case of self dual vacuum
Einstein equation, and hyperK\"ahler geometry also area preserving
diffeomorphism appear, where the spectral variable is merely
parameter. But in case of sDiff$(2)$ K.P. equation the situation
is different, where Takasaki--Takebe showed that one has to treat
$\lambda$ as a true variable and it enters into the definition of
the Poisson bracket. 
Our construction is closely related to Takasaki-Takebe
construction, only
the
K\"ahler like two form and the associated ``Darboux coordinates''
is replaced by volume form and the Poisson bracket is replaced by
its higher order Poisson bracket called Nambu bracket.

Suppose we consider $L = L(\lambda, p, q)$, $M = M(\lambda, p, q)$ and
$N = N(\lambda, p, q)$ are some Laurent series in $\lambda$ with coefficients are
functions of $p$ and $q$.

\begin{definition}\label{definition:2}
The volume preserving integrable hierarchy is defined by
\begin{gather}
\frac{\partial L}{\partial t_n} = \{B_{1n}, B_{2n}, L\},\label{guha:21}\\
\frac{\partial M}{\partial t_n} = \{B_{1n}, B_{2n}, M\}, \label{guha:22}\\
\frac{\partial N}{\partial t_n} = \{B_{1n}, B_{2n}, N \},\label{guha:23}
 \end{gather}
and
\begin{gather}
\{L, M, N \} = 1,\label{guha:24}
\end{gather}
where $ B_{1n} := (L^n)_{n \geq 0}$
and $ B_{2n} := (M^n)_{n \geq 0}$.
The first three equations are hierarchy equations, and these generates the
three flows of the system and the last one shows the volume preservation condition.
 The Nambu--Poisson bracket in 3D ``phase space''
 $(\lambda, p, q)$ is given by
\[
\{A(\lambda, p, q), B(\lambda, p, q), C(\lambda, p, q) \} =
\frac{\partial A}{\partial \lambda}\left(\frac{\partial B}{\partial p}
\frac{\partial C}{\partial q} - \frac{\partial B}{\partial q}
\frac{\partial C}{\partial p}\right) +  \mbox{cyclic   terms}.
\]
\end{definition}

Let us now compare our case with the area preserving KP hierarchy.
The sdiff(2) KP hierarchy is given by
\[ \frac{\partial {\cal L}}{\partial t_n} = \{ {\cal B}_n, {\cal L} \},
\qquad  \frac{\partial {\cal K}}{\partial t_n} = \{ {\cal B}_n, {\cal K} \},
\qquad  \{ {\cal L}, {\cal K} \} = 1,
\]
where ${\cal L}$ is a Laurent series in an indeterminant $\lambda$ of the form
\[
 {\cal L} = \lambda + \sum_{n=1}^{\infty} u_{n+1}(t) \lambda^{-n},
\]
$ {\cal B}_n = ({\cal L})_{\geq 0}$. The function ${\cal K}$ is called
Orlov function and it is defined by
\[
 {\cal K} = \sum_{n=1}^{\infty} nt_n {\cal L}^{n-1} + x + \sum_{i=1}^{\infty}
v_i {\cal L}^{-i-1},
\]
where $t_1 = x$.

\begin{remark}\label{remark:1}
Our hierarchy has a structure of volume preserving KP hierarchy
and instead of one Orlov function ${\cal K}$, we need two Orlov functions
$M$ and $N$.
\end{remark}

Any two equations of the hierarchy commute
\[
 \partial_{t_m}\partial_{t_l} = \partial_{t_l}\partial_{t_m},
\]
hence we obtain
\begin{proposition}\label{proposition:1}
The Lax equation for $L$, $M$, $N$ are equivalent to the following equations:
\begin{gather*}
\left\{ \frac{\partial B_{1n}}{\partial t_m},B_{2n} \right\} +
\left\{ B_{1n}, \frac{\partial B_{2n}}{\partial t_m}\right\}
 + \left\{ {\hat B_1}, B_{2m} \right\}\\
\qquad {} - \left\{ \frac{\partial B_{1m}}{\partial t_n}, B_{2m} \right\}
- \left\{ B_{1m}, \frac{\partial B_{2m}}{\partial t_n} \right\} +
\left\{ B_{1m}, {\hat H}_2 \right\}  = 0,
\end{gather*}
where
\[ {\hat H}_1 = \{ B_{1n}, B_{2n}, B_{1m} \} , \qquad
{\hat H}_2 =  \{B_{1n}, B_{2n}, B_{2m} \}.
\]
\end{proposition}

\begin{proof}
 Result follows from the compatibility conditions of hierarchy equations
 and the fundamental identity.
\end{proof}

\begin{remark}\label{remark:2}
In the case of sDiff$(2)$ hierarchy the above expression boils down to
zero curvature equation.
\end{remark}

Let $\Omega^{(3)}$ be a three form given by
\begin{definition}\label{definition:3}
\begin{equation}
 \Omega^{(3)} := \sum_{n=1}^{\infty} dB_{1n} \wedge dB_{2n} \wedge dt_{n} =
d\lambda \wedge dp \wedge dq + \sum_{n=2}^{\infty} dB_{1n} \wedge dB_{2n}
\wedge dt_{n}.\label{guha:25}
\end{equation}
\end{definition}

From the definition it is clear $\Omega$ is closed $3$ form.
In fact sDiff$(3)$ structure is clearly exhibited from this structure
and the theory is integrable in the sense of nonlinear graviton construction~[14].
This is a generalization of nonlinear graviton construction.

\begin{theorem}\label{theorem:2}
The volume preserving hierarchy is equivalent to the exterior differential equation
 \begin{equation} \Omega^{(3)}= dL \wedge dM \wedge dN.
\label{guha:26}
 \end{equation}
\end{theorem}

\begin{proof}
 We have seen that $\Omega$ can be written in two ways.
Expanding both sides of the exterior differential equation as linear
combinations of $ d\lambda \wedge dp \wedge dq $, $d\lambda \wedge dp \wedge dt_n$,
$d\lambda \wedge dq \wedge dt_n$ and $dp \wedge dq \wedge dt_n$.

When we pick up the coefficients of $ d\lambda \wedge dp \wedge dq$, we obtain
the volume preserving condition
\[
\{ L, M, N \} = 1.
\]

When we equate the other coefficients, viz. $ d\lambda \wedge dp \wedge dt_n$,
$d\lambda \wedge dq \wedge dt_n$ and $ dp \wedge dq \wedge dt_n$ we obtain
the following identities:
\begin{gather*} \frac{\partial ( B_{1n}, B_{2n} )}{\partial (\lambda , p)} =
\frac{\partial (L, M, N)}{\partial (\lambda, p, t_n)},\qquad 
\frac{\partial (B_{1n},B_{2n})}{\partial (\lambda, q)} =
\frac{\partial (L, M, N)}{\partial (\lambda, q, t_n )}, \label{guha:28}
\end{gather*}
and
\[
\frac{\partial (B_{1n}, B_{2n})}{\partial ( p, q)} =
\frac{\partial (L, M, N)}{\partial ( p, q, t_n )}\label{guha:29}
\]
respectively. We multiply the above three equations by $\frac{\partial L}{\partial q}$,
$ \frac{\partial L}{\partial p}$ and $ \frac{\partial L}{\partial \lambda}$ respectively.
If we add first and third equations and substract the second one from them,
then after using the volume preserving identity we obtain
\[
 \frac{\partial L}{\partial t_n} =  \{ B_{1n}, B_{2n}, L \}.
\]
Similarly we can obtain the other equations also, in that case we will
multiply the equations (\ref{guha:21}), (\ref{guha:22}) and
(\ref{guha:23}) by $\frac{\partial M}{\partial q}$,
$\frac{\partial M}{\partial p}$ and $\frac{\partial M}{\partial \lambda}$ respectively.
\end{proof}

Equation (\ref{guha:25}) and (\ref{guha:26}), we obtain
\[
d\left( M dL \wedge dN + \sum_{n=1}^{\infty} B_{1n} dB_{2n} \wedge dt_n\right) = 0.
\label{guha:30}
\]

This implies the existence of one form $Q$ such that
 \[
dQ = M d(LdN) + \sum_{n=1}^{\infty} B_{1n} d(B_{2n} dt_n).\label{guha:31}
 \]

This is an analogue of ``Krichever potential'' in the volume preserving case.
Hence we can say from (\ref{guha:24})
\begin{gather*}
 M = \frac{\partial Q}{\partial (LdN)}|_{B_{2n},t_n {\rm \  fixed }},\qquad 
B_{1n} = \frac{\partial Q}{\partial (B_{2n}dt_n)}|_{L,N,B_{2m},t_m ( m \neq n)
{\rm \  fixed}}.  \label{guha:33}
\end{gather*}

\section{Application to mulitidimesional integrable systems\\
and Riemann--Hilbert problem}

We already stated that our situation is quite similar to nonlinear graviton
construction of Penrose~[17] for the self dual Einstein equation. This is a
generalization of nonlinear graviton constructions.

To the geometer self dual gravity is nothing but Ricci flat K\"ahler geometry
and it is characterized by the underlying symmetry groups sDiff$(2)$, this are
called area preserving diffeomorphism group on surfaces. These are the natural
generalization of the groups Diff$(S^1)$, diffeomorphism of circle.

It is well known how the area preserving diffeomorphism group appears in the self
dual gravity equation. Let us give a very rapid description of this.

Let us start from a complexified metric of the following form
\[
 ds^2 = \det \left(\begin{array}{cc}
                             e^{11} &  e^{12} \\
                                     e^{21}  &  e^{22}
                                  \end{array}\right) = e^{11}e^{22} - e^{12}e^{21},
\]
where $e^{ij}$ are independent one forms. Ricci flatness condition boils down to
the closedness~of
\begin{equation} d\Omega^{kl} = 0 \label{guha:34}
\end{equation}
of the exterior $2$-forms
\[
\Omega^{kl} = \frac{1}{2} J_{ij} e^{ik} \wedge e^{jl}, \label{guha:35}
\]
where $J$ is the normalized symplectic form
\[
J = \left(\begin{array}{cc}
         0 & 1 \\
               -1 & 0
          \end{array}\right).
\]
Then above system of two
forms can be recast to
\[
 \Omega (\lambda ) = \frac{1}{2} J_{ij} \left( e^{i1} + e^{i2} \lambda \right)
\wedge \left( e^{j1} + e^{j2}\lambda \right).\label{guha:36}
\]
This satisfies
\begin{gather} \label{guha:37}
\Omega (\lambda ) \wedge \Omega (\lambda) = 0, \qquad
d \Omega (\lambda) = 0,
 \end{gather}
where $d$ stands for total differentiation. These suggest us to introduce a pair of
Darboux coordinates
\[
 \Omega (\lambda) = dP \wedge dQ
\]
and these are the sections of the twistor fibration
  \[
\pi : {\cal T} \longrightarrow CP^1,
\]
where ${\cal T}$ is the curved twistor space introduced by Penrose. Basically
each fibre is endowed with a symplectic form and as the base point moves this
also deforms and  here comes the area preservation.

Two pairs of Darboux coordiantes are related by
\begin{gather*}
  f (\lambda, P(\lambda), Q(\lambda) )= P^{\prime},\qquad
 g (\lambda, P(\lambda), Q(\lambda) ) = Q^{\prime}
\end{gather*}
and $f$ and $g$ satisfy $ \{ f , g \} =1.$ The pair $(f,g)$ is called
twistor data. Locally $f$ and $g$  (after twisting with $\lambda$) yield
patching function. Ricci flat K\"ahler metric is locally encoded in this data.
This set up is nothing but the Riemann--Hilbert problem in area preserving
diffeomorphism case.

The novelty of this approach is that this twistor construction will work in the
higher dimensions too, when there is no twistor projection. The most important
example is the electro-vacuum equation, volume preserving diffeomorphism groups
in three dimension play a vital role here. This model was first introduced
by Flaherty~[18] and later Takasaki~[2] showed how this works explicitly.

\subsection{Gindikin's bundle of forms}

We already stated that anti-self dual vacuum equations govern the behaviour of
complex 4-metrics  of signature $(+,+,-,-)$ whose Ricci curvature is zero and
whose Weyl curvature is self dual. These two curvatures are independent of change
of coordinates, so in one particular of the equations these metric becomes
autometically K\"ahler and can be expressed in terms of a single scalar function
$\Omega$, the K\"ahler potential. Then curvarure conditions will lead you to
Ist Plebenski's Heavenly equation
\[
 \frac{\p^2 {\tilde \Omega}}{{\partial x}{\partial \tilde x}} \frac{\p^2 {\tilde \Omega}}
{{\partial y}{\partial \tilde y}} - \frac{\p^2 {\tilde \Omega}}{{\partial x}{\partial \tilde y}}
\frac{\p^2 {\tilde \Omega}}{{\partial y}{\partial \tilde x}} = 1,\label{guha:39}
\]
or using the Poisson bracket (w.r.t $x$ and $y$)
\[
\{{\tilde \Omega}_{\tilde x}, {\tilde \Omega}_{\tilde y} \} = 1,
\]
and the correspoding anti-self-dual Ricci flat metric is
\[
g({\tilde \Omega}) = \frac{\p^2 {\tilde \Omega}}{{\partial x^i}{\partial \tilde x^j}} dx^i d{\tilde x}^j,
\qquad {\tilde x}^i = {\tilde x}, {\tilde y}, \qquad x^j = x, y.
\]

This system is completely integrable and it is an example of a multidimensional
integrable system.

Let $\Omega$ be the 2-form, given by
\begin{gather*}
\Omega (x,{\tilde x}, y, {\tilde y})
 = dx \wedge dy \\
\qquad
{}+ \lambda ({\tilde \Omega}_{x{\tilde x}}dx \wedge d{\tilde x} +
  {\tilde \Omega}_{x{\tilde y}}dx \wedge d{\tilde y} +
{\tilde \Omega}_{y{\tilde x}}dy \wedge d{\tilde x} +
{\tilde \Omega}_{y{\tilde y}}dy \wedge d{\tilde y}) +
  \lambda^2 d{\tilde x}\wedge d{\tilde y}.
\end{gather*}

\begin{lemma}\label{lemma:1}
\begin{gather*}
1) \quad  d\Omega = 0,\\
2) \quad  \Omega \wedge \Omega = 0.
\end{gather*}
\end{lemma}

\begin{proof}
 Since $\Omega$ satisfies Plebenski's Heavenly Ist equation, hence 2)
is true.
\end{proof}

A number of multidimensional integrable systems can be written in terms of a
2-form $\Omega$ which satisfies the equations (\ref{guha:34}).

\begin{example}
The dispersionless KP hierarchy has a Lax
representation with respect to a
series of independent (``time'') variables $ t= (t_1, t_2, \ldots )$
\[
\frac{\partial {\cal L}}{\partial t_n} = \{ B_n , {\cal L} \},
\]
where $ B_n := ({\cal L}^n)_{\geq 0}$,   $n= 1,2, \ldots$,
${\cal L}$ is a Laurent series in an indeterminant $\lambda$ of the form
\[
{\cal L}= \lambda + \sum_{n=1}^{\infty} u_{n+1}(t)\lambda^{-n},
\]
$\{~,~\}$ is a Poisson bracket in 2D phase space with respect to $(\lambda , x)$.

Let us consider
\[
\Omega = d\lambda \wedge dx + \sum_{n=2}^{\infty} dB_n \wedge dt_n
\]
then $ \Omega \wedge \Omega$ is equivalent to the zero curvature condition
\[
\frac{\partial B_n}{\partial t_m} - \frac{\partial B_m}{\partial t_n} +\{B_n , B_m \} = 0.
\label{guha:40}
\]
This is an alternative form of dispersionless KP hierarchy.

All these systems are related to area preserving diffeomorphism group sDiff(2).
\end{example}

In this paper we are presenting an analogous picture for multidimensional
integrable systems related to volume preserving diffeomorphism group.

\begin{proposition}\label{proposition:2}
 The 3-form
\[
\Omega^{(3)} = d\lambda \wedge dp \wedge dq + \sum_{n=2}^{\infty}dB_{1n} \wedge
dB_{2n} \wedge dt_n,
\]
 satisfies
\[
 d\Omega^{(3)} = 0, \qquad \Omega^{(3)} \wedge \Omega^{(3)} = 0.
\]
\end{proposition}

\begin{proof} This is equivalent to Propostion~\ref{proposition:1}, hence it is satisfied.
\end{proof}

The Gindikin's method [19] of pencil of 2-forms
is the  most effective way to study
these systems. Consider the following system of Ist order equation depending on a
parameter $ \tau = (\tau_1 ,\tau_2 ) \in {\mathbb C}^2 $
\begin{gather*}
 e^1 (\tau) = e^{11} \tau_{1}^{k} + \cdots + e^{1k}\tau_{2}^{k}, \\
 e^2 (\tau)= e^{21} \tau_{1}^{k} + \cdots + e^{2k}\tau_{2}^{k}, \\
\cdots\cdots\cdots \cdots \cdots \cdots\cdots\cdots \cdots\\
 e^{2l}(\tau) = e^{2l1}\tau_{1}^{k}+ \cdots + e^{2lk}\tau_{2}^{k},
\end{gather*}
where $e^{ij}$ are $1$-forms. Let $\Omega^k (\tau)$ be the bundle of 2-forms
\[
\Omega^k (\tau) = e^1 (\tau) \wedge e^2 (\tau) + \cdots + e^{2l-1}(\tau)
\wedge e^{2l}(\tau)
\]
satisfying the conditions
\[
 \left(\Omega^k \right)^{l+1} = 0, \qquad  \left(\Omega^k\right)^l \neq 0, \qquad d\Omega^k = 0.
\]
 The bundle of forms actually encodes the integrability of the original system.
 In the special case $l=1$, $k=2$, we recover the Ricci flat
 metric
\[
g= e^{11}e^{22} - e^{12}e^{21}.
\]

\subsection{Higher dimensional analogue of Gindikin's Pencil}

Let us consider the following system of Ist order equations depending
on parameter $ \tau = (\tau_1, \tau_2, \tau_3)  \in  {\mathbb C}^3$,
\begin{gather*}
e^1 (\tau ) = e^{11}\tau_1 + e^{12}\tau_2 + e^{13}\tau_3, \\
 e^2 (\tau )= e^{21}\tau_1 + e^{22}\tau_2 + e^{23}\tau_3, \\
 e^3 (\tau ) = e^{31}\tau_1 + e^{32}\tau_2 + e^{33}\tau_3.
\end{gather*}

Just like the previous situation these bundle of forms then encodes the
integrability of the original system. The metric defined here is the higher
order analogue of Ricci metric in two form case.

This metric is given by,
\[
g = e^{11}e^{22}e^{33} - e^{11}e^{32}e^{23} + e^{12}e^{31}e^{23}
 - e^{12}e^{21}e^{33} + e^{13}e^{21}e^{32} - e^{13}e^{31}e^{22}.
\]

\begin{remark}\label{remark:3}
Since 3-form $\Omega^{(3)}$ satisfies
\[
 d\Omega^{(3)} = 0, \qquad
 \Omega^{(3)} \wedge \Omega^{(3)} = 0,
\]
so it denotes the volume preserving multidimensional integrable systems. These
  systems can be described by Gindikin's bundle of multi forms, higher
dimensional analogue of nonlinear graviton.
\end{remark}

\subsection{Twistor description of volume preserving multidimensional\\
integrable systems}

The natural question would be to find out the analogous Riemann--Hilbert problem
in the volume preserving case. Our situation is very similar to electro-vacuum equation.
Let us consider two sets of solutions of hierarchy $( L, M, N)$ and
$({\hat L}, {\hat M}, {\hat N})$ with different analysity.
Then there exist an invertible functional relation between these
two sets of functions such that it satisfies
\[
{\hat L} = f_1 ( L, M, N),   \qquad  {\hat M} = f_2 ( L, M, N ),
\qquad   {\hat N}= f_3 (L, M, N),
\]
where $ f_1 = f_1 (\lambda, p, q)$, $ f_2 = f_2 (\lambda, p, q)$ and
$ f_3 = f_3 (\lambda, p, q )$ are arbitrary holomorphic functions defined
in a neighbourhood
of $\lambda = \infty$ except at $ \lambda = \infty$.

We assume $f_1, f_2, f_3 $ satisfy the canonical Nambu Poisson relation
\[
\{ f_1, f_2, f_3 \} = 1. \label{guha:41}
\]

This is a kind of Riemann--Hilbert problem related to three dimensional diffeomorphisms.
In this case sDiff$(3)$ symmetries is clear, in fact sDiff$(3)$ group acts
on $( f_1, f_2, f_3 )$, we can lift this action on $ (L, M, N)$
and $({\hat L}, {\hat M}, {\hat N})$ via Riemann--Hilbert fatorization.

\subsection{Application to hydrodynamic type systems}

There are certain kind of integrable systems, called hydrodynamic type,
naturally arise in gas dynamics, hydrodynamics, chemical kinematics and may other
situations, these are given by
\[
 u_{t}^{i} = v_{j}^{i}(u) u_{x}^{j}, \label{guha:42}
\]
where $v_{j}^{i}(u)$ is an arbitrary $ N \times N$ matrix function of
\[
u = \left(u^1, \ldots ,u^N\right), \qquad  u^i = u^i(x,t), \quad i=1, \ldots ,N.
\]

The Hamiltonian systems of hydrodynamic type systems considered above
have the form
\[
u_{t}^{i} = \{u^i,H\},\label{guha:43}
\]
where $H = \int h(u)dx$, is a functional of hydrodynamic type.
The Poisson bracket of these systems has the form
\[
\{ u^i(x),u^j(y) \} = g^{ij}(u) \delta_x (x-y) +
b_{k}^{ij}(u) u_{x}^k \delta (x-y),
\label{guha:44}
\]
called Dubrovin--Novikov type Poisson bracket. Dubrovin--Novikov showed if the
metric is non degenerate i.e. det $[g^{ij}] \neq 0$
then the above bracket yields a Poisson bracket provided $g^{ij}(u)$ is a metric of zero
Riemannian Curvature.

A large class of them can be described by the Lax form,
\[
 \psi_x = zA\psi , \qquad \psi_t= zB\psi.
\]

From the compatibility condition we obtain,
\[
 A_t= B_x , \qquad AB = BA.
\]
These set of equations can be easily recasted to
\[
 d\Omega = 0, \qquad \Omega \wedge \Omega = 0,
\]
where $\Omega$ is a closed one form.

Hence has also a twistorial description. The upshot of this section is that
Gindikin pencil of forms can be applied to large number of classes integrable
systems, including the volume preserving integrable systems we propose here.

\section{Conclusion}

In this paper we have shown that there are large class of integrable systems
can be obtained from Nambu--Poisson mechanics. They belong to the same family
of self dual Einstein or dispersionless KP type equations.
In some sense these
integrable systems are higher dimensional generalization of self dual
Einstein equation. Hence these systems are describable via Gindikin's
bundle of forms, or twistor method. Our integrable systems are volume
preserving. We hope that eventually such construction will find their
use in the construction of volume preserving KP and Toda type equation.

We have also discussed in this paper that Nambu--Poisson manifold is an
useful tool to study volume preserving integrable systems.
Recently in membrane theory physicists [20] have
found M-algebra from M-brane, these are related to Nambu--Poisson mechanics.

There are certain problems we have not discussed in this paper, viz. the
quantization of these volume preserving generalized multidimensional
integrable systems.
Presumably, the method of star product quantization [21, 22, 23] would
be the best way quantize these systems and
instead of binary star product we need triple star
product~[24].

\subsection*{Acknowledgement}

Author would like to record here the debt which the work
owes to the papers of K~Takasaki and T~Takebe and L~Takhtajan and to many
illuminating discussions with Professors K~Takasaki. He is also grateful
to Professors Dmitry Alekseevsky, Simon Gindikin, Warner Nahm and
Ian Strachan for valuable discussions.

\label{guha-lastpage}

\end{document}